\documentstyle[epsf,prd,aps,floats]{revtex}

\begin{document}

\draft

\input epsf \renewcommand{\topfraction}{0.8}
\twocolumn[\hsize\textwidth\columnwidth\hsize\csname
@twocolumnfalse\endcsname

\title{
Hadronic Decay of Late-Decaying Particles and
Big-Bang Nucleosynthesis
}

\author{Masahiro Kawasaki$^{(a)}$,
Kazunori Kohri$^{(b)}$ and
Takeo Moroi$^{(c)}$
}

\address{
$^{(a)}$Research Center for the Early Universe, 
Graduate School of Science,
University of Tokyo, Tokyo 113-0033, Japan\\
$^{(b)}$ Department of Earth and Space Science, Osaka University,
Osaka 560-0043, Japan\\
$^{(c)}$Department of Physics, Tohoku University, 
Sendai 980-8578, Japan
}

\maketitle

\begin{abstract}
    We study the big-bang nucleosynthesis (BBN) scenario with
    late-decaying exotic particles with lifetime longer than $\sim
    1$ sec.  With a late-decaying particle in the early universe,
    predictions of the standard BBN scenario can be significantly
    altered. Therefore, we derive constraints on its primordial
    abundance.   We pay particular attention to hadronic decay modes
    of such particles.   We see that the non-thermal production
    process of D, ${\rm ^{3}He}$ and ${\rm ^6Li}$ provides 
    a stringent upper bound on
    the primordial abundance of late-decaying particles with hadronic
    branching ratio.
\end{abstract}
{\small \vspace{-0.75cm}\hspace{7cm}RESCEU-04/04, 
OU-TAP-225, 
TU-710, 
astro-ph/0402490}

\vspace{3mm}

]

\renewcommand{\thefootnote}{\arabic{footnote}}

\newcommand{\D}{ {\rm D} }
\newcommand{\Dt}{\frac{D}{Dt}}
\newcommand{\Lag}{\cal L}
\newcommand{\Siga}{\sum_{\alpha}}
\newcommand{\abs}[1]{{\left|{#1}\right|}}
\newcommand{\anti}[1]{{\overline{#1}}}
\newcommand{\beseven}{{\rm ^7Be}}
\newcommand{\cm}{ {\rm cm} }
\newcommand{\chuukakko}[1]{\left\{{#1}\right\}}
\newcommand{\daikakko}[1]{{\left[{#1}\right]}}
\newcommand{\dpa}{ {d^{3}}\tilde{p}_{\alpha} }
\newcommand{\dva}{ {d^{3}}v_{\alpha} }
\newcommand{\erg}{ {\rm erg} }
\newcommand{\ev}{ {\rm eV} }
\newcommand{\fa}{f_{\alpha}}
\newcommand{\g}{ {\rm g} }
\newcommand{\gev}{ {\rm GeV} }
\newcommand{\hethree}{{\rm ^3He}}
\newcommand{\hefour}{{\rm ^4He}}
\newcommand{\kakko}[1]{{\left({#1}\right)}}
\newcommand{\kev}{ {\rm keV} }
\newcommand{\lisix}{{\rm ^6Li}}
\newcommand{\liseven}{{\rm ^7Li}}
\newcommand{\mG}{ {M_{\rm G}} }
\newcommand{\ma}{m_{\alpha}}
\newcommand{\mb}{ {\rm mb} }
\newcommand{\mev}{ {\rm MeV} }
\newcommand{\mgrav}{ {m_{3/2}} }
\newcommand{\mmodulino}{ {m_{\tilde{\phi}}} }
\newcommand{\mphi}{ {m_{\phi}} }
\newcommand{\mpi}{ {m_{\pi}} }
\newcommand{\order}{{\cal O}}
\newcommand{\pa}{ {{p}_{\alpha}}}
\newcommand{\pbg}{ {{p}_{\rm bg}}}
\newcommand{\qa}{ {q_{\alpha}} }
\newcommand{\s}{ {\rm s} }
\newcommand{\staave}[1]{{\langle{#1}\rangle}}
\newcommand{\T}{ {\rm T} }
\newcommand{\tenten}{ {\cdot\cdot\cdot} }
\newcommand{\tev}{ {\rm TeV} }
\renewcommand{\thefootnote}{\#\arabic{footnote}}
\newcommand{\vect}[1]{{\mbox{\boldmath {$#1$}}}}
\newcommand{\va}{ {v_{\alpha}} }
\newcommand{\Y}{ {\rm Y}}
\newcommand{\yp}{ {Y}_{\rm p}}
\newcommand{\y}{ {\rm Y}}

\setcounter{footnote}{0}

It has long been recognized that the (standard)
big-bang nucleosynthesis (BBN)  provide  a good probe for the early
universe.  With our current knowledges of nuclear reaction processes,
we can precisely calculate abundances of the light elements (in
particular, ${\rm D}$, ${\rm ^3He}$, ${\rm ^4He}$, ${\rm ^6Li}$, and
${\rm ^7Li}$) as functions of the baryon-to-photon ratio $\eta\equiv
n_{B}/n_\gamma$.  Thus, comparing the theoretical predictions with the
observations, we can obtain various informations about the evolution
of the universe.  Importantly, using the value of $\eta$ suggested by
the WMAP ($\eta =(6.1 \pm 0.3 )\times 10^{-10}$
\cite{Spergel:2003cb}), the theoretical predictions show relatively
good agreement with the observations.

In cosmological scenarios in the frameworks of  physics beyond the
standard model, however, the BBN may not proceed in the standard way.
This is because, if we assume physics beyond the standard model, there
exist various exotic particles.  Those exotic particles may cause
non-standard processes and spoil the success of the standard BBN.

In particular, if the exotic particle (called $X$ hereafter) decays
radiatively and/or hadronically after the BBN starts, the primordial
abundances of the light elements may be significantly affected.
Indeed, energetic particles produced by the decay of $X$ may scatter
off and dissociate the background nuclei.  If such processes occur
with sizable rates, predictions of the standard BBN scenario are
changed.

In various models of particle physics, there exist long-lived (but
unstable) particles and hence their effects on the BBN should be
studied.  In particular, many of those particles interact very weakly
and hence it is difficult to study their properties by collider
experiments.  Thus, in some case, the BBN provides useful and
important informations about those weakly interacting particles.
Probably the most famous example of such late-decaying particle is the
gravitino in the supergravity theory.  Gravitinos are produced in the
very early universe by scattering processes of the particles in the
thermal bath.  Their interaction is suppressed by the inverse powers
of the (reduced) Planck scale $M_*$, and hence the lifetime becomes
very long.  An order of magnitude estimate shows that, if the
gravitino is lighter than $\sim 10\ {\rm TeV}$, its lifetime becomes
longer than $\sim 1\ {\rm sec}$.  In this case, the thermally produced
gravitinos decay after the BBN starts.

Effects of the radiative decay of such long-lived particles have been
extensively studied (see, e.g.,
\cite{Ellis:1990nb,KawMor,Jedamzik:1999di,Cyburt:2002uv}).  However,
in many cases, hadronic branching ratio may not be negligible and
hence, in studying the BBN with late-decaying particles, it is
necessary to consider effects of the hadronic decay processes.  Even
if $X$ dominantly decays into the photon (and something else),
hadronic branching ratio is expected to be at least $10^{-(2-3)}$
since the emitted photon can be converted to a quark-antiquark pair.
Of course, if $X$ directly interacts with the colored particles,
hadronic branching ratio may become larger.

In the past, the BBN with hadro-dissociation processes induced by
hadronic decays of long-lived particles was studied in
\cite{HadronicDecay}, which are effective for relatively long lifetime
($\gtrsim 10^{2}$ sec).\footnote
{BBN constraints from the interconversion process between neutrons and
protons by hadronic decays were studied in
Refs.~\cite{Reno:1987qw,Kohri:2001jx}, which is effective for shorter
lifetime ($\lesssim 10^{2}$ sec).}
The analysis in Ref.~\cite{HadronicDecay} contains, however, a lot of
room to be improved since many of nuclear reactions that they used
were not accurate enough or not available at that time.  After the
study of \cite{HadronicDecay}, however, there have been significant
theoretical, experimental and observational progresses in the study of
the BBN.  First of all, new data for the hadron reactions have become
available and their qualities have been improved very much. Moreover,
the primordial abundances of the light elements have been precisely
determined with various new observations.  In addition, it has been
recently known that some of the non-standard processes induced by the
decay of late-decaying particle $X$, which were not taken into account
in \cite{HadronicDecay}, may play important roles in the BBN.  With
these progresses, a new study of the late-decaying particles with
hadronic branching ratio should be relevant.

Thus, in this letter, we reconsider the BBN processes with long-lived
exotic particle $X$ paying particular attention to the effects of the
hadronic decay modes.  As a result, we will see that, with  hadronic
decay modes, the constraint on the primordial abundance of $X$ becomes
very severe compared to the case only with the radiative decay modes.
In particular, we will see that non-thermal production of D, ${\rm ^{3}He}$ and ${\rm ^6Li}$ provides a stringent constraint.

We first introduce the framework of our study.  Although we have
several candidates of the late-decaying particles, we perform
our analysis as model-independently as possible.  Thus we parameterize
the property of the late-decaying particle $X$ using the following
parameters: $\epsilon_X$ (released energy from the single decay of
$X$) which is equal to its mass $m_X$ unless specially stated,
$E_{\rm jet}$ (energy of the primary parton from the decay of $X$),
$\tau_X$ (lifetime), $B_{\rm h}$ (hadronic branching ratio),
and the primordial abundance of $X$.
We parameterize the primordial abundance by using the following
``yield variable'' $Y_X\equiv n_X/s$, which is defined at the cosmic
time $t\ll \tau_X$.  Here, $n_X$ is the number density of $X$ while
$s$ is the total entropy density.
We assume that $X$ decays only into the particle in the ``observable
sector,'' and that the branching ratio decaying into the hidden-sector
particle vanishes.

In our analysis, we calculate the primordial abundances of the light
elements for given sets of the parameters listed above taking account
of dissociation processes due to hadronic (and electromagnetic)
interactions.  The outline of our calculation is as follows.  In order
to study the effects of the hadronic decay modes, we first calculate
the energy spectrum of the (primary) hadrons (in particular, protons
and neutrons) generated from the partons directly emitted from $X$.
Then, we follow evolutions of the hadronic showers.  As a result, the
numbers of the light elements produced (or destroyed) by the decay of
$X$ is calculated.  We include the hadro-dissociation processes (as
well as the photo-dissociation ones) into the BBN reaction network and
numerically follow the evolution of the abundances of the light
elements.  Details of our study will be described elsewhere
\cite{KKM_Preparation}.

When $X$ decays hadronically, quarks or gluons (i.e., partons) are
first emitted.  Those partons are hadronized soon after the
decay. Thus, we should consider how the nucleons (and the mesons)
propagate in the thermal bath.  

In considering the propagation of stable particles (i.e., proton,
neutron, and heavier nuclei), there are two classes of important
processes.  The first is the scattering process with the background
photons $\gamma_{\rm BG}$s and electrons $e_{\rm BG}^-$s.  By
scattering off $\gamma_{\rm BG}$s and $e_{\rm BG}^-$s, energetic
nuclei lose their energy without affecting the abundances of light
elements.  The second class is the scatterings with the background
nuclei.  With such processes, first of all, the background nuclei
become energetic after the scattering. Therefore the energetic nuclei
are copiously produced.  (We call this ``hadronic shower.'')  In
addition, if inelastic scattering occurs, background nuclei are
dissociated and the abundances of the light elements are changed.
Thus, if the second class of reactions occur significantly, the
abundances of the light elements deviate from the predictions of the
standard BBN.

Including relevant hadronic scattering processes (as well as
photo-dissociation processes), we have calculated the abundances of
the light elements.  In our study of the evolution of the hadronic
showers, the basic framework is the same as that used in
\cite{HadronicDecay} although there are several modifications.  The
most important improvements are as follows. (i) We carefully take into
account the energy loss processes for high-energy nuclei through the
scattering with background photons or electrons. In particular,
dependence on the cosmic temperature, the initial energies of nuclei,
and the background $\hefour$ abundance are considered. (ii) We adopt
all the available data of cross sections and transfered energies of
elastic and inelastic hadron-hadron scattering processes. (iii) The
time evolution of the energy distribution functions of high-energy
nuclei are computed with proper energy resolution. (iv) The JETSET 7.4
Monte Carlo event generator~\cite{Sjostrand:1994yb} is used to obtain
the initial spectrum of hadrons produced by the decay of $X$. (v) The
most resent data of observational light element abundances are
adopted. (vi) We estimate uncertainties with Monte Carlo simulation
which includes the experimental errors of the cross sections and
transfered energies, and uncertainty of the 
baryon to photon ratio $\eta$. 
(We take $\eta = (6.1\pm 0.3)\times 10^{-10}$.)

We are interested in the situation where the number density of $X$ is
small enough so that the energetic particles in the hadronic shower
dominantly scatter off the background particles. (Otherwise, the
abundances of the light elements are so affected that the results
become inconsistent with the observations.)   If the energetic nuclei
lose most of their energy by scattering off $\gamma_{\rm BG}$s and
$e_{\rm BG}^-$s, the number densities of the (background) nuclei is
almost unaffected.   On the contrary, if the scattering rate with
$\gamma_{\rm BG}$s and $e_{\rm BG}^-$s becomes negligible, number
densities of the light elements are significantly changed by the
hadronic processes.

For charged nuclei, energy-loss rates due to the scatterings with the
$\gamma_{\rm BG}$ and $e_{\rm BG}^-$ depend on the velocity of the
nuclei \cite{Reno:1987qw}.  For the temperature $T\lesssim 20$ ${\rm
keV}$, relativistic nuclei do not lose their energy by the scatterings
with $\gamma_{\rm BG}$s and $e_{\rm BG}^-$s.  For non-relativistic
nuclei with velosity $\beta_N$, energy-loss is dominated by the
scattering process with background electrons with velosity
$\beta_e<\beta_N$.  (As pointed out in \cite{Reno:1987qw}, energy loss
via the scatterings with high-velosity electron is extremely
suppressed and is negligible in our case.)  Thus, we use the
energy-loss rate
\begin{eqnarray}
  \frac{dE_N}{dt} &=& 
  -\frac{4\pi\alpha^{2}Z^{2}n_e}{m_{e}\beta_N} 
  I(\beta_N/\sqrt{2T/m_e}) 
  \ln\left(\frac{\Lambda m_e \beta_N^2}{\omega_p}\right),
  \label{eq:energyloss}
\end{eqnarray}
where $\omega_{\rm p}$ is the plasma frequency, $m_{e}$ is the
electron mass, $Z$ is the charge of the nuclei, $\alpha$ is the fine
structure constant, $\Lambda$ is a constant of $O(1)$.  (In our
numerical calculations, we take $\Lambda =1$.)  In addition,
\begin{eqnarray}
  I (r) = \frac{4}{\sqrt{\pi}} \int_0^r dx x^2 e^{-x^2}.
\end{eqnarray}
Notice that the number density of the background electron with
$\beta_e<\beta_N$ is given by $n_eI(\beta_N/\sqrt{2T/m_e})$.  For
$\beta_N\gg\sqrt{T/m_e}$, $I\simeq 1$ and the energy-loss is very
efficient.  On the contrary, once the velosity of the nuclei becomes
smaller than the thermal velosity of the electron
$\langle\beta_e\rangle\sim\sqrt{T/m_e}$, $I\sim
O((\beta_N/\sqrt{T/m_e})^3)$ and energy loss becomes less
effective.\footnote
{In the earlier version of this letter, we did not take into account
the velocity distribution of $e_{\rm BG}^-$ and used the formula for
the case when all the background electrons have the same velosity
$\langle\beta_e\rangle$.  Thus, for $\beta_N<\langle\beta_e\rangle$,
the energy-loss rate we used was incorrect and was underestimated.
This resulted in an overestimate of the non-thermally produced ${\rm
^{6}Li}$.}

For the neutral particle (i.e., neutron), on the contrary,
the scattering process with $e_{\rm BG}^-$ is not important for
$T\lesssim 0.1\ {\rm MeV}$.  Thus, once energetic neutrons are
produced, they may induce hadronic showers by scattering off the
background nuclei and change their abundances.  If the temperature
becomes lower than  $\sim 0.3 \ {\rm keV}$, however, most of the
(energetic) neutrons decay into protons  before scattering off the
background particles.  Thus, if $\tau_X\gtrsim 3 \times 10^{7}  \ {\rm
sec}$,  the most important constraint on $Y_X$ is from the
photo-dissociation processes of the light elements.

In our analysis, we also take account of the photo-dissociation
processes induced by high-energy photons generated from the decay
products of $X$.  Once $X$ decays, energy as large as $(1-B_{h})
\epsilon_X$ is directly deposited into the radiation while the rest
of the released energy $B_{h}\epsilon_X$ first goes into the
hadronic sector.  Importantly, however, spectrum of the high energy
photon primarily depends on the total amount of the injected energy
(as well as the temperature).  Thus, we approximate that all the
emitted energy in the decay process is eventually converted to the
form of radiation, calculate the photon spectrum, and estimate the
photo-dissociation rates of the light elements.  For details of the
treatment of the radiatively decaying particles, see \cite{KawMor}.

Moreover, pions and  kaons produced  in the hadronization are also
considered.  For $T\gtrsim  0.1$~MeV ($t \lesssim 10^{2}$~sec) they
can scatter  off the background nuclei before their decay, which leads
to  increase the neutron to proton ratio and hence the abundance of D
and $^{4}$He. In this letter, we have conservatively omitted the
similar effects induced by $n \bar{n}$ and $p \bar{p}$
pairs  which are
effective for $t \gtrsim 10^{2}$~sec~\cite{Kohri:2001jx} because of
insufficient data for the cross sections of energetic $\bar{n}$ and
$\bar{p}$. If we include them, the constraint can become
severer~\cite{Kohri:2001jx}.

With the procedure explained above, we have calculated the abundances
of the light elements for a given set of the model parameters.  By
comparing the results with the observations, we derived upper bounds
on $Y_X$.  As observational constraints on the light element
abundances, we adopt the following values, D/H = $(2.8 \pm 0.4) \times
10^{-5}$~\cite{Kirkman:2003uv}, $\hefour$ mass fraction $\yp = 0.238
\pm 0.002 \pm 0.005 $ by Fields and Olive (FO)~\cite{Fields:1998} and
$\yp = 0.242 \pm 0.002 (\pm 0.005)_{\rm syst}$ by Izotov and Thuan
(IT)~\cite{Izotov:2003xn}, $\log_{10}(\liseven/{\rm H}) = -9.66 \pm
0.056 (\pm 0.3)_{\rm syst}$~\cite{Bonifacio:2002}, $\lisix/\liseven <
0.07 (2\sigma)$~\cite{li6_obs}, and $\hethree/\D < 1.13
(2\sigma)$~\cite{Geiss93}.  The above errors are at 1 $\sigma$ level
unless otherwise stated. Here we added the systematic error to $\yp$
(IT) which is as same as the value in Fields and Olive.  For
$\liseven$/H we also added the additional systematic error for the
possibilities that $\liseven$ in halo stars might have been depleted
in stars or supplemented by production in cosmic-ray
interactions~\cite{Fields:1996yw,Hagiwara:fs}. For $\liseven/{\rm H}$
we conservatively use only the upper bound since the experimental data
for the non-thermal ${\rm ^7Li}$ production by energetic $\hefour$,
especially for the transfered energy to $\hefour$ through the
inelastic collisions between nucleon and $\hefour$, are insufficient,
by which we do not include them in our computation.

\begin{figure}[t!]
    \begin{center}
        \epsfxsize=0.45\textwidth\epsfbox{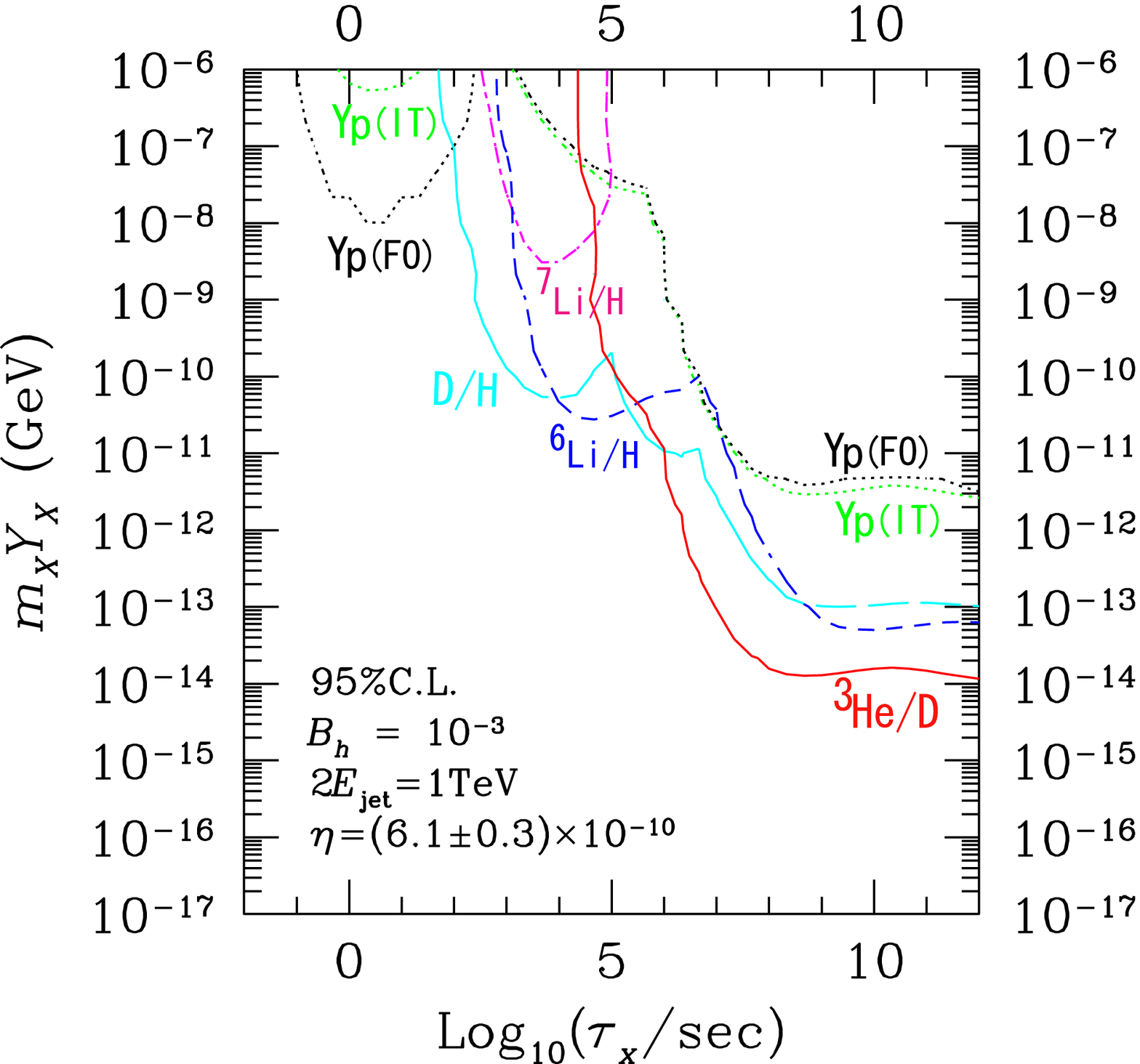}
        \caption{Upper bounds on $m_X Y_X$ as a function of $\tau_X$
        at 95\% C.L.  for the case of $B_{\rm h}=10^{-3}$. The name of
        the element which gives the constraint is written by each
        line. We assume that two hadron jets are produced by single
        decay of $X$ with the energy $ E_{\rm jet} = m_{X}/2$. Here we
        consider $m_{X}=\epsilon_{X}=1\ {\rm TeV}$. Note that $Y_X =
        n_{X}/s$.}
        \label{fig:SmallBh}
        \epsfxsize=0.45\textwidth\epsfbox{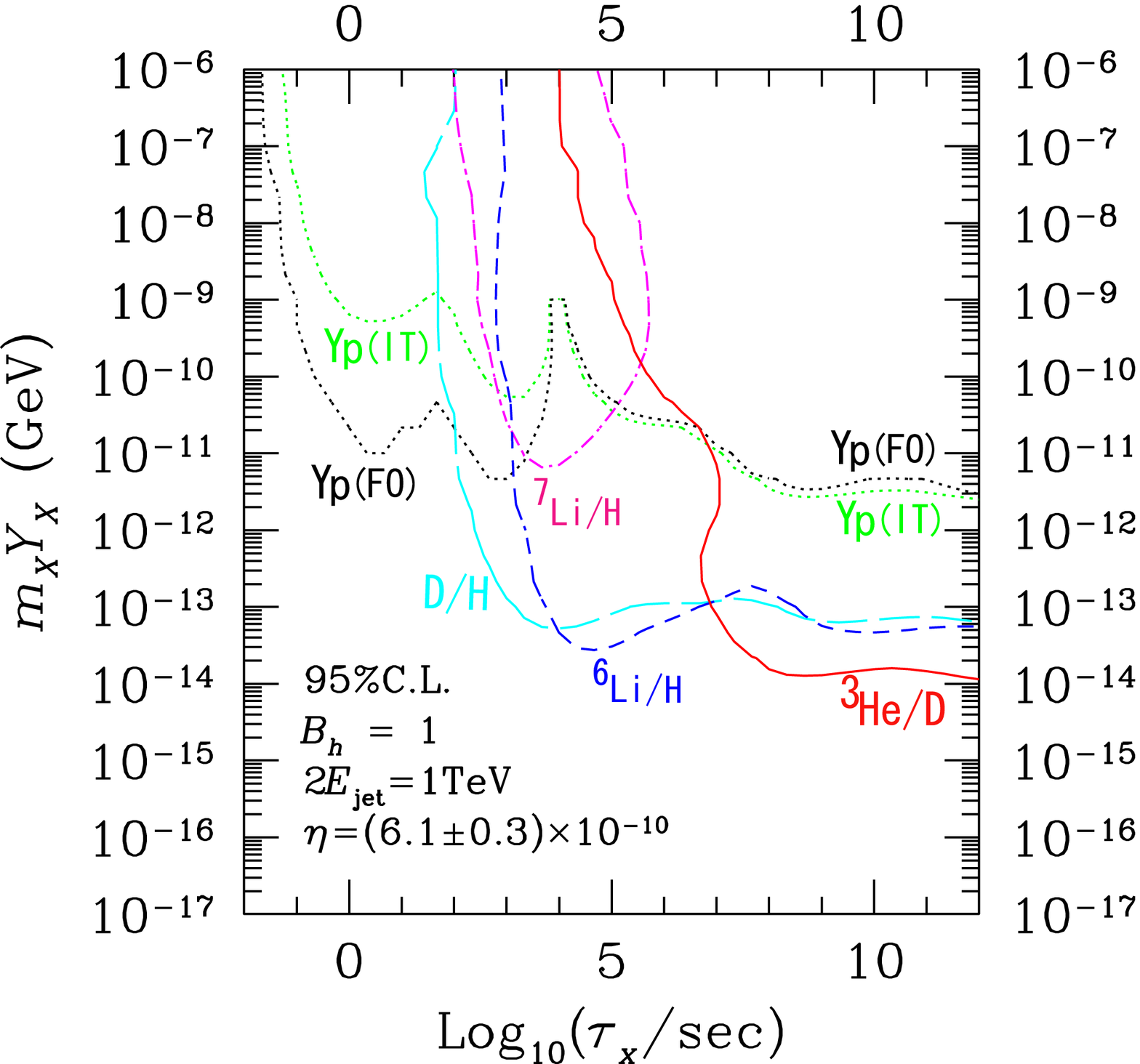}
        \caption{Same as Fig.\ \protect\ref{fig:SmallBh}, except for
        $B_{\rm h}=1$. }
        \label{fig:LargeBh}
    \end{center}
\vspace{-0.5cm}
\end{figure}

In Figs.\ \ref{fig:SmallBh} and \ref{fig:LargeBh}, we show the upper
bound on $m_XY_X$ as a function of the lifetime $\tau_X$ for $B_{\rm
h}=10^{-3}$ and $1$.  (Here, we used $m_{X}=1\ {\rm TeV}$ although,
even if we vary $m_{X}$ within 0.1 $-$ 100 TeV, the bounds do not
change significantly \cite{KKM_Preparation}.)  In deriving the bound
we estimate the confidence levels by the $\chi^2$ fitting including
both the theoretical and the observational errors.  As we mentioned,
for the case with very long lifetime, upper bound is almost the same
as the one for the case with radiatively decaying particles
\cite{KawMor}. The most stringent constraint comes from the
overproduction of ${\rm ^{3}He}$.\footnote{
However, even for $\tau_{X}\gtrsim 10^{8}$~sec,  
the hadronic processes give some contribution to the D production, 
which can be seen from  comparison between Fig.~\ref{fig:SmallBh} 
and Fig.~\ref{fig:LargeBh}.
}

For shorter lifetime $\tau_X\sim 10^{2-8}\ {\rm sec}$, 
overproductions
of ${\rm D}$ and ${\rm ^{6}Li}$ provide constraints on $m_XY_X$.
${\rm D}$ is mainly produced by the hadro-dissociation of ${\rm
^4He}$.  Notice that the hadro-dissociation of ${\rm ^4He}$ is
possible with energetic $p$ and $n$ even at such early epoch although
the photo-dissociation of ${\rm ^4He}$ can be effective only at $t
\gtrsim 10^{6}\ {\rm sec}$ ($T \lesssim 1$~keV) since high energy
photons lose their energy by scatterings off the background photons
before interacting with ${\rm ^4He}$ at earlier epoch.  
The overproduction of ${\rm ^6Li}$ is 
mostly due to the non-thermal processes
with energetic ${\rm T}$ and ${\rm ^3He}$ which are generated by the
hadro-dissociation of ${\rm ^4He}$.  These ${\rm T}$ and ${\rm ^3He}$
can be sufficiently energetic and may scatter off the background ${\rm
^4He}$ to produce ${\rm ^6Li}$ via the processes ${\rm T}+{\rm
^4He}\rightarrow {\rm ^6Li}+n$ and ${\rm ^3He}+{\rm ^4He}\rightarrow
{\rm ^6Li}+p$.

{One might think that the produced ${\rm ^6Li}$ may be energetic and
scatter off background nuclei to be destroyed.  In our numerical
calculations, we obtain the energy distribution of the non-thermally
produced ${\rm ^6Li}$ and calculated the surviving rate of such ${\rm
^6Li}$.  We have found that, for $\tau_X\gtrsim 100\ {\rm sec}$, the
surviving rate is almost $1$.  Therefore, one does not have to take
account of the destruction of the produced ${\rm ^6Li}$.  If the
temperature is higher than $\sim 10\ {\rm keV}$, however,
non-thermally produced ${\rm ^6Li}$ is destructed by the thermal
process ${\rm ^6Li}(p, {\rm ^4He}){\rm ^3He}$, as was also pointed out 
in \cite{Jedamzik:2004ip}.  Thus the constraint from the 
${\rm ^6Li}$ is weakened for $\tau_X\gtrsim 10^4\ {\rm sec}$.}

If the lifetime of $X$ is short (i.e., $\tau_X\lesssim 10^2\ {\rm
sec}$), $X$ decays when the number density of the background electron
(and positron) is still abundant and, in this case, energetic nucleons
and nuclei are likely to scatter off the background electron and lose
their energy without committing the hadro-dissociation processes.  In
this case, the production of D and ${\rm ^6Li}$ is suppressed and the
upper bound on $m_XY_X$ is not stringent.  

Finally, we apply the above results to the primordial abundance of
gravitinos.  In the inflationary universe, gravitinos are produced by
the scattering processes of the thermal particles.  The yield variable
of the gravitino is proportional to the reheating temperature $T_{\rm
R}$ after the inflation, $Y_X= 1.5 \times 10^{-12}\times (T_{\rm
R}/10^{10}\ {\rm GeV})$ \cite{KawMor}.  In addition, assuming the
(massless) gauge boson and gaugino as the final state, lifetime of the
gravitino is given by $\tau_{3/2}\simeq 4\times 10^8\ {\rm sec}\times
N_{\rm G}^{-1}(m_{3/2}/100 \ {\rm GeV})^{-3}$, where $N_{\rm G}$ is
the number of the generators of the gauge group, and $m_{3/2}$ is the
gravitino mass.  As examples, we consider two typical cases.  The one
is the case where the gravitino dominantly decays into the photon and
photino, producing two hadron jets with $E_{\rm
jet}=\frac{1}{3}m_{3/2}$; in this case, we take $B_{\rm h}=10^{-3}$,
$N_{\rm G}=1$, and $\epsilon_X=\frac{1}{2}m_{3/2}$.  The other is the
case where the gravitino dominantly decay into the gluon and gluino,
producing one hadron jet with $E_{\rm jet}=\frac{1}{2}m_{3/2}$; in
this case we take $B_{\rm h}=1$, $N_{\rm G}=8$, and
$\epsilon_X=\frac{1}{2}m_{3/2}$.  For these cases, we read off the
upper bound on the reheating temperature for several values of the
gravitino mass.  The results are shown in Table \ref{table:tr}. It is
seen that the constraint on $T_{\rm R}$ is much more stringent than
that obtained for gravitino without hadronic decay.

\begin{table}
    \begin{center}
        \begin{tabular}{lccc}
            & $B_{\rm h}=10^{-3}$ & $B_{\rm h}=1$ & \\
            \hline
            $m_{3/2} = 100\ {\rm GeV}$ &
            $2 \times 10^{6}\ {\rm GeV}$ &
            $3 \times 10^{6 }\ {\rm GeV}$ & \\
            $m_{3/2} = 300\ {\rm GeV}$ &
            $3 \times 10^{ 6}\ {\rm GeV}$ &
            $3 \times 10^{6 }\ {\rm GeV}$ & \\
            $m_{3/2} = 1\ {\rm TeV}$ &
            $3 \times 10^{8 }\ {\rm GeV}$ &
            $3 \times 10^{5 }\ {\rm GeV}$ & \\
            $m_{3/2} = 3\ {\rm TeV}$ &
            $2 \times 10^{8 }\ {\rm GeV}$ &
            $7 \times 10^{5 }\ {\rm GeV}$ & \\
        \end{tabular}
        \caption{Upper bounds on $T_{\rm R}$ for several values of
        $m_{3/2}$.  }
        \label{table:tr}
    \end{center}
\vspace{-1cm}
\end{table}

\noindent
{\it Note added:} While finalizing this letter, we found the paper by
K. Jedamzik~\cite{Jedamzik:2004er} which have some overlap with our
analysis.

\noindent
{\it Acknowledgements:} This work was partially  supported by the
Grant-in-Aid for Scientific Research from the Ministry of Education,
Science, Sports, and Culture of Japan, No.~15-03605 (KK) and
No.~15540247(TM).

\end{document}